\newcommand{\be}{\begin{equation}}
\newcommand{\ee}{\end{equation}}
\newcommand{\bee}{\begin{eqnarray}}
\newcommand{\eee}{\end{eqnarray}}

\newcommand{\Intg}{\int_{-\infty}^\infty}

\documentstyle[twocolumn,prb,aps]{revtex} 

\begin{document}
\wideabs{ 
\title{Comment on ``Exact non-equlibrium transport
through point contacts \\ 
in quantum wires and fractional quantum Hall devices''}
\author{Sergei Skorik}
\address{Department of Physics, Ben-Gurion University, Beer-Sheva 84105,
Israel \\
and Department of Physics, Weizmann Institute for Science, Rehovot 76100,
Israel}
\date{20 August 1997}
\maketitle
\begin{abstract}
Fendley, Ludwig and Saleur (\prb {\bf 52}, 8934 (1995)) have obtained
an expression for the non-equlibrium current through a constriction
in the quantum Hall bar
based on the Bethe ansatz technique and the Bolzmann equation.
In this Comment we draw attention to a serious flaw in their derivation.
We argue that their result  is correct in the linear
response limit but should be taken with a care out of equilibrium
for finite bias. The reason is in the use of the equilibrium
scattering matrix in the place of the transition probability amplitude out
of equilibrium, tbe substitution which was not shown to be legimate.  
\end{abstract}
} 

Two years ago, Fendley Ludwig and Saleur, \cite{FLS}   
based on the standard Bethe ansatz technique, obtained
a non-perturbative expression for the backscattering 
current out of equilibrium through a single impurity
in the quantum Hall bar with $\nu=\case{1}{3}$. The principal
novel  component of their method is
the ``fusion'' of thermodynamic Bethe ansatz with the Bolzmann rate equation,
implemented as follows.
The quantum Hall bar with impurity, described by the Luttinger
theory with point-like backscattering term $\lambda_{bs}
\delta(x)(\Psi_L^+\Psi_R+\Psi_R^+\Psi_L)$ 
was mapped onto the boundary sine-Gordon model 
\be
H_{BSG}={1\over 2}\int_0^\infty\!\!\!\! dx[\Pi^2(x)+(\partial_x\Phi)^2]+
\lambda_{bs}\cos{\sqrt{8\pi\nu}\over 2}\Phi(0), \label{BSG}
\ee
which is exactly solvable in equilibrium
and can be diagonalized by the Bethe ansatz.
Transport of Laughlin $(e/3)$ quasiparticles through impurity 
 maps onto  
a scattering of sine-Gordon quasiparticles (kinks, antikinks and breathers)
off the boundary. All the interaction in (\ref{BSG}) is now at the boundary,
which behaves like a non-elastic scatterer: kink can be bounced as 
an antikink and vice versa, which changes the charge of the system. 
However, it is
a bare Hamiltonian where the bulk interactions are absent. As a result
of diagonalization,
sine-Gordon quasiparticles interact also in the bulk by a pairwise
point-like interaction that adds merely a phase-shift with momentum
of each quasiparticle preserved -- a consequence of the peculiar 
conservation
laws of (\ref{BSG}), and the distribution function of the gas of
quasiparticles differs from the usual Fermi one.  
Under these circumstances, the Bolzmann rate equation was employed
to obtain the current:
\be 
I_B={e\over h}\int_0^\infty \!\!{dp\over p}[n_+(p,\mu,T)-n_-(p,\mu,T)]
|S_+^-(p,\lambda_{bs})|^2.
\label{Bolz}
\ee
This expression was referred in literature to as an {\it exact} result,
since  the exact quantum expressions are known for all the entries
of the integral in (\ref{Bolz}) in equilibrium.
In particular, the kink-antikink scattering matrix
element $S^-_+$ was obtained in Ref.\ \onlinecite{Zamo}, 
while the density of states
of quasiparticles $n_{\pm}$ were found from the thermodynamic Bethe ansatz
(TBA) \cite{FLS,AlZamo}.

We agree with the authors of Ref.\  \onlinecite{FLS} that,
although in the absense of microscopic derivation
 not an exact quantum equation for the current,
Bolzmann rate equation in its probabilistic interpretation 
and with the exact quantum entries could give an exact result
when applied to the rather ``idealistic'' quasi-particles of the sine-Gordon
model with their peculiar scattering properties. We want to draw
the reader's attention, however, to the quantities itself employed
by Fendley, Ludwig and Saleur as the entries of the Bolzmann equation,
namely,
the equilibrium bulk densities $n_{\pm}$ in the presense of bias
but in the absense of backscattering, and the equilibrium scattering
matrices $S^-_+$ for zero bias. The approximation of the densities of states
of quasiparticles by their equilibrium values seems  to be legimate
for a point-like impurity in the limit when size $L\to\infty$. However,
use of the equilibrium  impurity scattering matrices in the place of
the transition probability out of equilibrium, $W=\langle+|-\rangle$,
 seems to be unjustified
and lacks support in Ref.\  \onlinecite{FLS}. The quantity $W$ out
of equilibrium, when properly calculated,
 does not in general necessarily coincide with the equilibrium
scattering matrix, while in Ref.\  \onlinecite{FLS} these two have been 
tacitly identified, and thus a strong conjecture has been put forward
that needs to be proved.
In other words, keeping
finite $\mu$ in the bulk density of states $n_{\pm}$ while using
equilibrium scattering matrix in (\ref{Bolz})
is beyond the allowed accuracy and the whole expression might 
have no physical sense, unless by a proper 
calculation it can be derived rigorously.

In order to illustrate our point, we will resort to a few examples.   

\paragraph{} The first example is the Kondo model in a magnetic
field, which is closely related to the boundary sine-Gordon model.
Magnetic field here plays a role of the bias, although one is in the
equilibrium situation. Magnetoresistance, calculated by N.Andrei
for the Kondo model by means of Bethe ansatz \cite{Andrei}, is
expressed through the elastic scattering matrix in the linear response
limit \cite{Lang}. 
Since the ground state is unstable with respect to switching
on a magnetic field, the scattering matrix turns out to depend
on the structure of the ground state and, hence, on the magnetic field.  

\paragraph{} The second example is the Anderson model \cite{ANDR}
out of equilibrium. For this model an exact expression for the quantum
current was derived by Wingreen and Meir: \cite{Meir}
\be
I={2e\Gamma\over h}\Intg d\omega[f_L-f_R]\text{Im}G^R_{dd}(\omega),
\label{eq:I}
\ee
where $f_{L,R}=f(\omega-\mu_{L,R})$ are the Fermi functions, $\Gamma$
is the hopping strength from the impurity to the leads analogous
to $\lambda_{bs}^2$ and $G^R_{dd}$ is the impurity retarded
Green function.  Equations for
current (\ref{Bolz}) and (\ref{eq:I}) bear formal resemblence.
It seems natural that in the case of quantum Hall bar, when the
leads consist of the interacting Luttinger liquid, Fermi-functions
of Eq.\ (\ref{eq:I}) are substituted by the distribution
functions of the Luttinger liquid, $n_{\pm}$, in Eq.\ (\ref{Bolz}).
More important difference appears in the use of the non-equilibrium
spectral function in the exact quantum current instead of the 
equilibrium scattering matrix. Indeed, as can be easily checked by the Keldysh
technique for the Anderson model, \cite{Meir2,Meir3,Hersh2}
 the non-equilibrium impurity spectral density
$\text{Im}\, G^R$ {\it is in general 
voltage-dependent}, whereas in Ref.\ \onlinecite{FLS} the authors 
employed the {\it equilibrium} scattering matrices of 
Ref.\ \onlinecite{Zamo} for the
transition probability. Is the appropriate quantity for the quantum
Hall bar indeed voltage-independent? We see no intuitive grounds for 
ruling out voltage-dependence {\it ab initio}.
Indeed, although the impurity in the quantum Hall bar
is represented in the Hamiltonian as a simple potential scattering
and has no dynamical degrees of freedom as opposed to the Anderson case,
the renormilized impurity is ``dressed'' by the interactions in the leads
and becomes rather non-trivial,
so that the voltage-independence would be possible only due
to some rather special circumstances and requires theoretical confirmation.  
It appears that support given in Ref.\ \onlinecite{FLS} is not enough:
checking against the exact result in the $\nu=\case{1}{2}$ case
is not sufficient. Both in the Andreson model and in the
quantum Hall bar the retarded Green function and the scattering matrix
are not expected to depend on voltage in the absense of interactions
(for vanishing on-site interactions $G^R_{dd}=(\omega-\epsilon_0+i\Gamma)^{-1}
$, while scattering over the ground state filled by non-interacting
particles leaves the ground state undisturbed and is not sensitive
to the chemical potential). So, the voltage-dependence appears to
be a result of interactions. 

Another supporting evidence for (\ref{Bolz}) given in Ref.\ \onlinecite{FLS}
is the linear response limit $V\to 0$, checked against the Keldysh
calculation. However, in this limit
one is allowed to substitute the equilibrium transition probability
into  (\ref{Bolz}) and the agreement is expected, too. 

Finally, one may pose a question whether there exists
a well-defined analogue of the non-equilibrium transition probability
in the framework of the standard Bethe ansatz. Heuristically,
since the ground state of (\ref{BSG}) is unstable with respect 
to switching on a bias, the
correct voltage-dependent transition probabilities  
 can be obtained when one considers the scattering
of kinks over the voltage-dependent ground state filled by anti-kinks,
rather than over an empty vacuum state. However, it seems to
be impossible to perform such a calculation in practice. The sourse
of the difficulty is that for the boundary sine-Gordon 
model an additional bias term $\mu\Phi(x=0)$ has not
been shown to be compatible with the integrability of (\ref{BSG})
and, therefore, an additional parameter $\mu$ cannot be accomodated by
the S-mtraices of Ref.\ \onlinecite{Zamo}. 
The authors of Ref.\ \onlinecite{FLS} have tacitly avoided
this problem by obtaining separately scattering matrices for zero bias,
and bulk densities for non-zero bias but without an impurity,
thus bringing together quantities from two different problems.

In conclusion, we argued that the result of Fendley, Ludwig and Saleur
for the current based on the Eq.\ (\ref{Bolz}) is correct in the linear
response limit but should be taken with a care out of equilibrium
for finite bias. The reason is in the use of the equilibrium
scattering matrix in the place of the transition probability amplitude out
of equilibrium, tbe substitution which was not shown to be legimate.  
The additional support, e.\ g. a direct microscopic derivation
of Eq.\ (\ref{Bolz}) out of equilibrium is necessary.  

I am  grateful to
 Natan Andrei and Yigal Meir for the  very helpful discussions.
This work was supported
 by Kreitman and by Kaufmann fellowships, and partially funded
by the Israel Science Foundation, grant No 3/96-1.




   
\end{document}